\documentclass[fleqn,10pt]{wlscirep}
\usepackage[utf8]{inputenc}
\usepackage[T1]{fontenc}
\usepackage{lineno}
\linenumbers

\title{EuroCrops: All you need to know about the Largest Harmonised Open Crop Dataset Across the European Union}

\author[1,*]{Maja Schneider}
\author[1]{Tobias Schelte}
\author[1]{Felix Schmitz}
\author[1]{Marco K\"orner}
\affil[1]{Technical University of Munich (TUM), TUM School of Engineering and Design, Munich, 80333, Germany}

\affil[*]{corresponding author: Maja Schneider (maja.schneider@tum.de)}

\usepackage[
  acronym,
]{glossaries-extra}

\setabbreviationstyle[acronym]{long-short}
\defglsdisplayfirst{\emph{#1#4}}
\defglsdisplayfirst[acronym]{\emph{#1#4}}
\glsdisablehyper

\newcommand\blfootnote[1]{%
    \begingroup
    \renewcommand\thefootnote{}\footnote{#1}%
    \addtocounter{footnote}{-1}%
    \endgroup
}

\newacronym{gl:hcat}{HCAT}{hierarchical crop and agriculture taxonomy}
\newacronym{gl:hcatv2}{HCATv2}{hierarchical crop and agriculture taxonomy -- version 2}
\newacronym{gl:cap}{CAP}{common agricultural policy}
\newacronym{gl:eu}{EU}{European Union}
\newacronym{gl:ec}{EC}{European Commission}
\newacronym{gl:un}{UN}{United Nations}
\newacronym{gl:wfs}{WFS}{web feature service}
\newacronym{gl:gpkg}{GPKG}{GeoPackage}
\newacronym{gl:gis}{GIS}{geographic information system}
\newacronym{gl:jrc}{JRC}{Joint Research Centre}
\newacronym{gl:eo}{EO}{Earth observation}
\newacronym{gl:sdg}{SDG}{sustainability development goal}
\newacronym{gl:doi}{DOI}{Digital Object Identifier}

\newglossaryentry{gl:apia}{name=Agenţia de Plăţi şi Intervenţie pentru Agricultură, description={}}

\begin{abstract}
  EuroCrops contains geo-referenced polygons of agricultural croplands from 16 countries of the \gls{gl:eu} as well as information on the respective crop species grown there.   
  These semantic annotations are derived from self-declarations by farmers receiving subsidies under the \gls{gl:cap} of the \gls{gl:ec}.  
  Over the last 1.5 years, the individual national crop datasets have been manually collected, the crop classes have been translated into the English language and transferred into the newly developed \gls{gl:hcat}. 
  
  EuroCrops is publicly available under continuous improvement through an active user community.
  
  \glsresetall
\end{abstract}

\begin{document}
\nolinenumbers

\flushbottom
\maketitle

\thispagestyle{empty}


\section{Background \& Summary}

As\blfootnote{Preprint. Under review.} the world's population continues to grow and global climate change becomes increasingly apparent, enhancing the efficiency and resilience of agriculture at both the local and global level is a crucial challenge for humanity's future.
Recent developments in satellite-based \gls{gl:eo} have provided us with the ability to observe and analyse the processes occurring on the Earth's surface in near real-time.
By leveraging machine learning and artificial intelligence, we can extract valuable insights from these enormous volumes of high-quality and information-rich data, which can inform the development of functional process models for the monitoring of agricultural crops and the design of future applications.
For example, the activity of these vegetation stands could be monitored and deviations from the expected progression, and thus the expected crop yields, could be detected.
Based on this information, farmers would be able to initiate countermeasures at an early stage.
This would make a decisive contribution to food security, representing one of the central \glspl{gl:sdg} stated by the \gls{gl:un}.
However, these possibilities are massively limited by the insufficient availability of qualitative reference data, which are necessary for the creation of functional process models on the basis of such Earth observation data.

The EuroCrops project aims to show how this gap can be filled by compiling administrative data assessed in the context of agricultural subsidy control in the \gls{gl:eu} area, exemplified by a first pilot project.\cite{schneider2021epe}
In light of previous studies, e.g., BreizhCrops \cite{russwurm2020breizhcrops}, ZueriCrop \cite{turkoglu2021crop}, and CropHarvest \cite{tseng2021cropharvest}, the key objectives of EuroCrops lie in the extension of both the variability of crop species classes to be represented and the geographical scale of the considered regions.

For this purpose, we collected geo-referenced crop datasets from three countries within Europe, harmonised the data by translating the crop names and developed an hierarchical structure to order the occurring crops.
Finally, the crop labels were paired with the corresponding Sentinel-2 \gls{gl:eo} data and we released the TinyEuroCrops\cite{schneider2021tinyeurocrops} dataset publicly via the repository of the Technical University of Munich.
Despite faced with some challenges, we soon realised that the dataset gained its popularity not due to the satellite data, but due to the fact that we also published the geo-referenced field polygon vector data together with the harmonised information of which crop species were cultivated there for a certain year.
Having this data prepared in one reconciled format, language, and centrally available across borders and not just on a national level sparked the discussions about its broad applicability in various domains.
The fact that this data has been prepared in a joint standardised format and language and that it is centrally available across borders and not only at national level has triggered discussion about its broad applicability in various areas.
The research questions related to the analysis of agricultural diversity and food security in Europe were one of the reasons for the popularity of the dataset, leading to the motivation to extend them spatially and later also temporally.

In this article, we present and describe the first spatially extended EuroCrops vector dataset.
For this release, we manually collected the raw crop declaration data from 16 \gls{gl:eu} countries, which was made available and distributed across multiple platforms and servers.
After translating the textual declarations data, we developed a new version of our \gls{gl:hcat}\cite{schneider2021epe} in order to organize all crops that are cultivated within the \gls{gl:eu} into a common hierarchical representation scheme.
The process of this development is visualised in Figure \ref{fig:process} and will be further explained in the methods section.

By being able to analyse agricultural data at this expanded spatial scale, which extends from Sweden to Portugal, we hope to enable researchers to carry out their work across borders and gain new insights.
We further publish the sources and links to the original datasets\cite{schneider2022githubwiki} and provide mappings\cite{schneider2022githubmappings} from the national crop class representation into \gls{gl:hcat} on our GitHub\cite{schneider2022github} repository.
This way, it is now possible to extend the dataset in time to the desired year by linking the mapping to a national dataset that can be found on the respective websites.

\section*{Methods}
\label{sec:methods}

\begin{figure}[ht]
  \centering
  \includegraphics[width=0.7\linewidth]{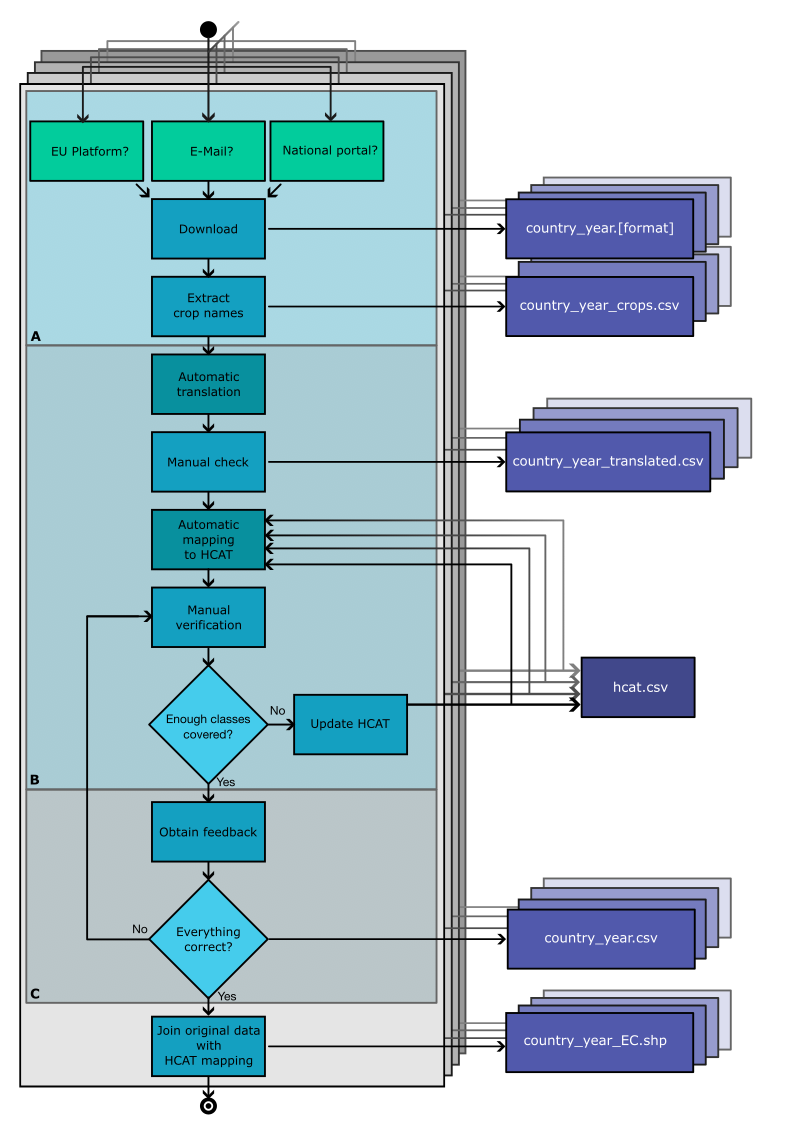}
  \caption{%
      The process of constructing the EuroCrops dataset.
      Each layer represents the process for one country with the three stages of development: \textbf{A. Data Collection}, \textbf{B. Harmonisation}, \textbf{C. Validation and Feedback}.
      Each stage has one or more outputs, indicated in purple, and with \texttt{country} and \texttt{year} being replaced accordingly.
      Only the \texttt{hcat.csv} exists once across all country-specific processes and gets gradually updated in each harmonisation step.
      While automatic sections exist, a manual check is required each time, making the progress in total heavily dependent on work that has to be done by hand.
  }
  \label{fig:process}
\end{figure}

In order to compile the presented dataset, several iterative steps had to be performed, which can roughly be grouped into \textit{data collection}, \textit{harmonisation} and  \textit{validation}, denoted as \textbf{A}, \textbf{B} and \textbf{C} in Figure \ref{fig:process} and will be further described in the next subsections.
Data obtained from each member state of the \gls{gl:eu} has to undergo the entire procedure, sometimes even multiple times, as indicated by the stacked layers in Figure \ref{fig:process} and arrows going from each countries \texttt{Update HCAT} process back to the beginning and the \texttt{Automatic mapping to HCAT} for the individual dataset.
This recurring loop is the main reason for the exponentially increasing amount of manual work that was necessary for the creation of the dataset and required careful deliberation on the right moment for cutting the development of \gls{gl:hcat}.

\subsection*{A. Data Collection}

\begin{figure}[ht]
    \centering
    \includegraphics[width=0.5\linewidth]{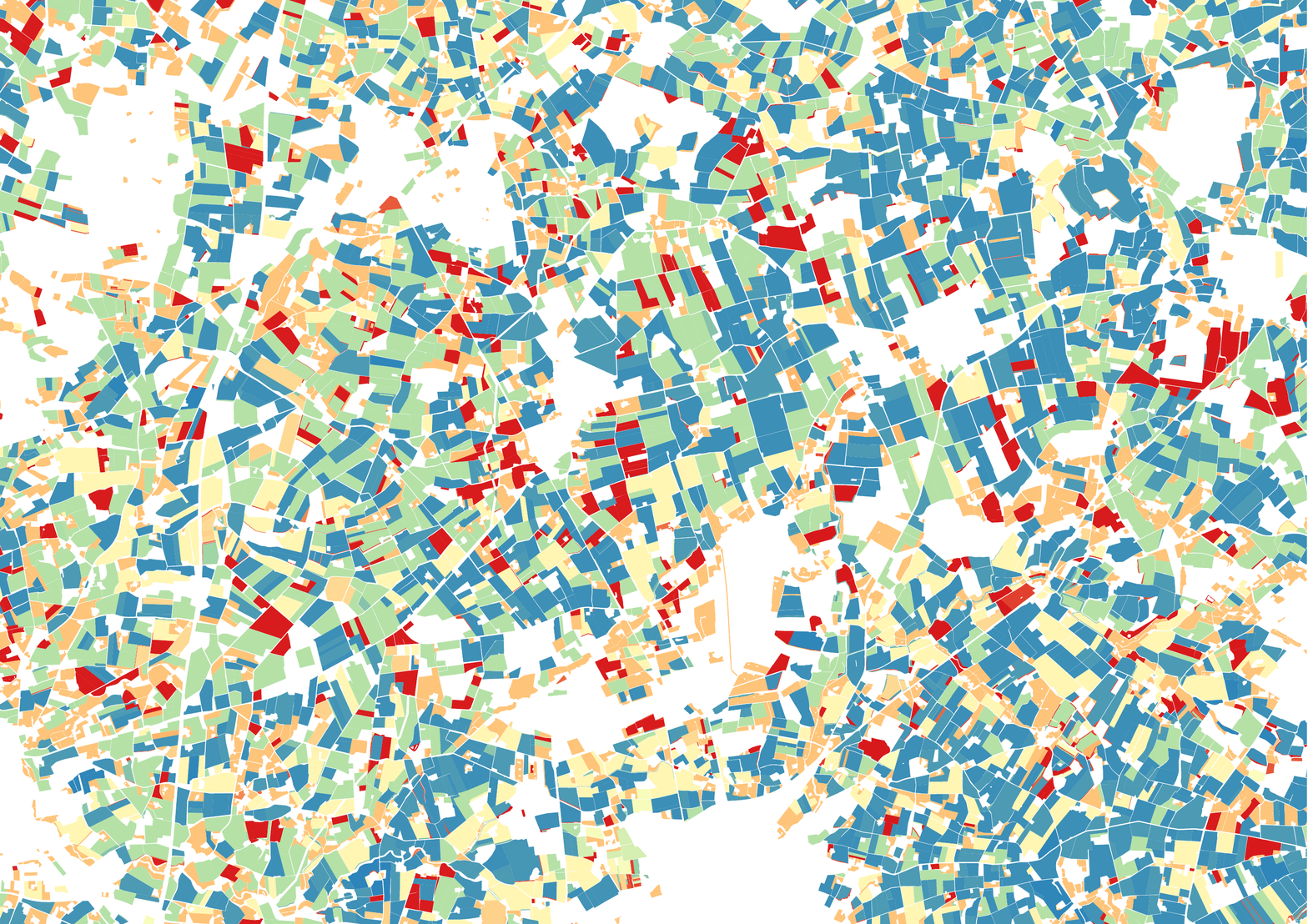}
    \caption{
        Exemplified raw input data with the corresponding attribute table shown in Table \ref{tab:raw_data}.
        This selections shows parts of the North Rhine-Westphalian (Germany) dataset\cite{denrw2021data} with each crop class being coloured differently.
        The data consists of geo-referenced polygons which indicate the field borders and hold information about the grown crop for a certain year.
}
    \label{fig:nrw}
\end{figure}

\begin{figure}[ht]
    \centering
    \includegraphics[width=1\linewidth]{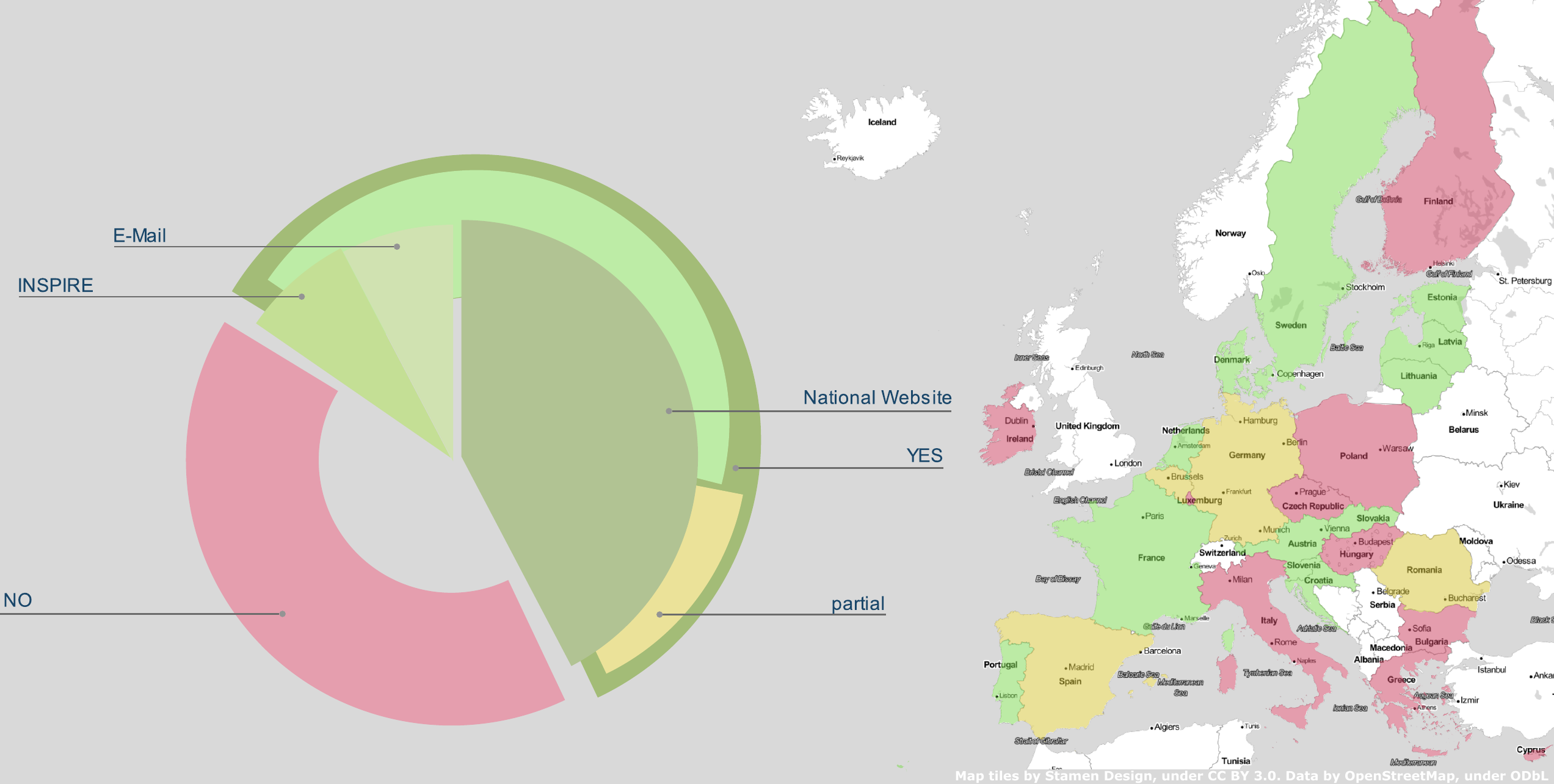}
    \caption{
        This diagram shows the data availability, coverage and access across the \gls{gl:eu}.
        The map on the right indicates whether data from a certain country is part of EuroCrops.
        Green indicates a full country participation, yellow are the member states with only partial coverage and data from red countries was not available at the time of the development of the dataset.
        Finland for instance released the data later and is still covered in the section "Data Records", but is not part of EuroCrops.
        On the left, the pie chart breaks down several analyses:
        Generally, red and green shades tell what fraction of data in the \gls{gl:eu} is available, while indicating with bright green and yellow full or partial coverage, as already on the map.
        The green pie segments on top visualise where the data was originally obtained from.
        This shows, similar to table \ref{tab:ministries} that the majority of the data originates from national websites, including geoportals.
}
    \label{fig:diagrams}
\end{figure}

As EuroCrops consists of multiple smaller datasets, the data collection itself plays an integral role.
This paper will focus on the practical part of that process, whereas an in-depth analysis of the challenges of creating a transnational dataset is described in more detail by Schneider et al. \cite{schneider2022challenges}.\\
Generally, we identified four ways of data acquisition:
Firstly, many countries publish national crop data on the webpage of the respective ministry or agency responsible for agricultural, food or rural topics.
Some countries instead offer a national geoportal, distributing different kinds of geodata specifically or, as another mean of distribution, publishing geodata on an international level, e.g. via \texttt{INSPIRE}\cite{inspire2022inspire} or \texttt{data.europe.eu}\cite{commission2022data}.
Lastly, if the data is not openly distributed on a webpage or geoportal, we reached out personally to ministries or agencies and asked for the data directly.
Most of the national datasets used in the EuroCrops project were collected from national ministry webpages or geoportals as listed in Tables \ref{tab:ministries} and \ref{tab:geoportal} respectively, mostly made available as \emph{ESRI shapefiles}, \texttt{GeoJSON}, or \texttt{GeoPackage} \texttt{(GPKG)}.
Nonetheless, some data can only be accessed via a \gls{gl:wfs} implemented in a \gls{gl:gis}, allowing the user to display the desired data and save it in a chosen file format.
The other means of data access are shown in Tables \ref{tab:direct} and \ref{tab:international}.
Figure \ref{fig:diagrams} puts all this information into context, gives an overview of the available datasets, and indicates from where the data originates.
Countries marked yellow in Figure \ref{fig:diagrams} indicate only partial availability of crop data for the respective country.
In order to give a better understanding of the original raw datasets we got from the countries, we visualised a small fraction of the data from North-Rhine Westphalia (Germany) in Figure \ref{fig:nrw} with coloured geo-referenced agricultural parcel polygons.
Table \ref{tab:raw_data} gives an impression of how the corresponding original raw attribute table looks like.
Each row entry describes the crop species that has been cultivated on the associated parcel.

\subsection*{B. Harmonising Country-Specific Crop Classes}

After collecting the data and extracting the set of country-specific crop classes from the attribute tables, we initiated the harmonisation process.
This step is necessary because the crop names from each country usually come in the national language of the member states and without standardised codes, as shown in Table \ref{fig:nrw}.\\
Instead of working with the entire attribute table, we worked with a table showing the name and code of a certain crop class per row together with its absolute and relative occurrence in the dataset.
In Figure \ref{fig:process}, that file, preserving original crop class name and code, is denoted as \texttt{country\_year\_crops.csv}.\\
Following this, the \texttt{automatic translation} starts the process of harmonising the given original crop class names into the \gls{gl:hcat} taxonomy.
Therefore, multiple steps had to be performed:
The file \texttt{country\_year\_translated.csv} arises from the translation of the crop classes into English.
Despite the access to modern translation programmes, we were not able to automate this part end-to-end, as country-specific agricultural terms seem to cause mistranslations across all common translators.
By correcting the translations manually, we hope to bridge that gap and make the dataset as reliable as possible.
Similarly, mapping the translations to \gls{gl:hcat} was only possible to perform automatically to a certain degree and required \texttt{manual checks} (see Figure \ref{fig:process}) as well.
This was again caused by the diversity of the crop classes declared by the member states.\\
Within this process of manual translation and matching, we were able to catch most of the missing classes in our growing taxonomy.
The iterative \texttt{updates of HCAT} helped in the detailed classification of delivered datafiles by the countries, but also shed light on relevant and focus areas within the taxonomy.
To the matching \gls{gl:hcat} name, we also added the corresponding \gls{gl:hcat} code, which embeds the hierarchy of the taxonomy.
This way, we enriched the country-specific original crop name and code with our \gls{gl:hcat} name and code and the absolute and relative occurrence in a country.\\
Hence, we are able to visualise the number of instances of certain crop classes and compare the occurrences with those from other countries for general diversity analysis and taxonomy class updates.
The preliminary file is stored in a \texttt{country\_year.csv} after positive assessment during working step \textbf{C}.

\subsection*{C. Community Work: Content Validation and Feedback Incorporation}

The largest expertise on country-specific crop classes still lies with the respective countries, driving to the decision to keep them onboard during the validation phase of the project.
Therefore, we asked all countries during the end phase of our pipeline if our translations and mappings seemed reasonable.
Out of 16 countries, we received feedback from seven who double-checked and reviewed our work.
While this increased the quality of the dataset, it also started another loop in the harmonisation block, which is visualised in Figure \ref{fig:process} as the arrow going from \texttt{Everything correct?} to \texttt{Manual verification}.
Eventually, we uploaded the first version of the dataset on our university-owned data-sharing platform and set up a GitHub repository\cite{schneider2022github} for the community to have a first look.
This resulted into several opened issues and pull requests where improvements to the mappings were suggested.
Each time we were content with a version of the mapping, we manually joined the original dataset with our mapping and saved it as a shapefile.
This lead to one shapefile for each country and five successive versions of the dataset incorporating the proposed changes from GitHub.
One exemplary attribute table of such a shapefile is shown in Table \ref{tab:ec_data}.
All of the versions were individually uploaded to Zenodo\cite{schneider2022ecz}, which now officially tracks the versions with a \gls{gl:doi}.

%

%
%
%
%

\section*{Data Records}

In the following paragraphs, all individual data sources are presented.
For each contributing country the data source, available years, coverage, licence, and format are described and referenced.
By doing so, we aspire to give the research community a tool to discover and access raw data faster and more reliably.

\paragraph{Austria}
The dataset for Austria comprises a vast range of years, spanning from 2015 to 2021.
Moreover, the whole territory of the country is covered without any regional omissions.
Crop classes are defined very detailed with an approximate number of 200 classes.
The files were made available in \texttt{GPKG} format via two platforms, the European “data.europa.eu”\cite{at2021data} and "data.gv.at"\cite{at2021data2}, a platform that distributes data of the public sector in Austria for further analysis and development.
However, both platforms receive the datasets from “Agrarmarkt Austria”, which is a public geodata office.
As such, data is published free of charge under the Creative Commons Licence CC-BY-AT 4.0.
In the course of the EuroCrops project, the dataset of 2021 was harmonised for Austria.

\paragraph{Belgium} 
Due to the federal structure of Belgium, the data is split into two sets covering the regions of Flanders and Wallonia.
Not only is the data published via different platforms, its structure also differs heavily between the two regions.\\
The data for Flanders\cite{befl2021data} is published by the Department of Agriculture and Fishery on its website as shapefiles.
Additionally, a word document explaining the current state of the data as well as the abbreviations that occur in the attribute table of the shapefiles is available in Flemish language.
The crop classes are differentiated very precisely with an approximate number of 275 classes.
Datasets are available for the years 2019, 2020 and 2021.\\
The datasets for Wallonia are published by the Geoportal of Wallonia\cite{bew2016data} as shapefiles.
With an approximate number of 150 classes the crop classification of Wallonia is still quite precise, even though the Flemish data is more detailed.
On the other hand, a wider time period is captured by the Wallonian datasets, covering all years since 2015.\\
So far only the Flemish data for the year 2021 got harmonised in the course of the EuroCrops-Project.

\paragraph{Croatia}
The Croatian data \cite{hr2020data} is distributed in \texttt{GPKG} format via a platform managed by the Agency for Payments in Agriculture, Fisheries and Rural Development, where an abundant sequence of years is available ranging from 2011 to 2021.
Due to translation difficulties, we obtained the data directly from the Paying Agency, and while all regions of the country are covered by the dataset, its differentiation between 14 crop classes turns out to be rather coarse.
For EuroCrops, the data of 2020 was harmonised.

\paragraph{Denmark}
The dataset of Denmark \cite{dk2019data} comprises of only the mainland.
The Faroe Islands and Greenland are not included.
However, with an approximate number of 300 classes, the Danish crop taxonomy is very detailed.
Datasets are available since the year 2017.
The data is available as shapefiles provided by the Danish Agricultural Agency.
All the data provided is considered open data, which means it can be openly used and distributed.
The Danish data of 2019 was harmonised throughout the course of the EuroCrops Project.

\paragraph{Estonia}
The Estonian dataset \cite{ee2021data} is made available under the creative commons licence.
Thus, there are no limitations to public access.
It can be acquired via the INSPIRE Geoportal as \gls{gl:wfs}.
When accessing the data via a \gls{gl:wfs} URL in a \gls{gl:gis}, the dataset can be transformed and saved as \texttt{GeoJSON} for example.
It covers all of Estonia but only for the current year.
Thus, data from 2021 was harmonised.
However, the crop differentiation is very precise, leading to a high number of ca. 150 classes.

\paragraph{Finland}
The Finnish dataset \cite{fi2021data} covers all provinces of the country.
Data is available for the years 2020 and 2021.
However, none of the years got harmonised yet, as Finland provided its datasets very late after the harmonisation process was already completed.
The data differentiates between 200 classes roughly, which enables a very precise crop classification.
The Finish Food Authority distributes the data via a \gls{gl:wfs} under the Creative Commons Licence BY 4.0.
Consequently, the datasets were implemented into a \gls{gl:gis} and saved as shapefile.

\paragraph{France}
France publishes national geodata as open licence on the "data.gouv.fr" platform\cite{fr2018data_official} as \texttt{GPKG}- and shapefiles.
While the central point of distribution makes it easy to discover and access the data, the fact that each region has its own sub-dataset makes the platform barely usable for someone who needs the entirety of the French data.
Luckily, there is a second (unofficial) server\cite{fr2018data_cq} that hosts a combination of all these national datasets in shapefile format.
Additionally, an excel sheet is available, containing the descriptions of all crop abbreviations used in the datasets.
The class differentiation is moderate.
Approximately 70 crop classes are distinguished.
In the course of the project, datasets were downloaded for the years spanning from 2016 to 2019, of which the file for 2018 was harmonised.
The data covers not only the French mainland but also overseas territories.

\paragraph{Germany}
Due to the federal structure of Germany datasets are not published on a national level, but by each federal state ("Bundesland") individually.
Two datasets were acquired:
One covers Lower Saxony \cite{dels2021data} and another one North Rhine-Westphalia \cite{denrw2021data}.
Both datasets depict the crop situation of 2021 and have a very high class precision, distinguishing between ca. 240 crop classes. Both files are distributed as shapefiles, one on the online platform for Rural Development and Agricultural Promotion of Lower Saxony, the other one on the geoportal of North Rhine-Westphalia.
Both datasets are published under “data licence Germany - attribution - Version 2.0”.
For the sake of completeness, it is worth noting that Brandenburg also published its data\cite{debb2021data}, but has not been included into EuroCrops yet.

\paragraph{Latvia}
The Rural Support Service of Latvia provides a \gls{gl:wfs}, which can be used to implement and convert the Latvian files to \texttt{GeoJSON} or shapefile in a \gls{gl:gis}.
The data \cite{lv2021data} is open so there are no publishing restrictions.
The files cover the whole territory of the country and are available for 2021 and 2022.
The file for 2021 got harmonised in the course of the EuroCrops Project.
The class precision is very high, differentiating between 150 crop types approximately.

\paragraph{Lithuania}
The crop parcels of Lithuania \cite{lt2021data} are available as shapefiles for the year 2021 covering the whole territory of the country.
Consequently, data got harmonised for the aforementioned year.
The file differentiates between 24 crop classes only.
However, the chosen classes are precise.
Datasets of a similar low number of classes normally assign very general crop terms to the classes (i.e. vineyard, citrus fruits, grassland).
In the case of Lithuania, the crop types assigned to the classes are very specific.
Thus, the class precision can be defined as medium, despite its low number of actual classes.
The data is published via Geoportal.lt, a platform distributing Lithuanian geo-data as part of the INSPIRE directive.

\paragraph{Netherlands}
The Dutch Ministry of Economic Affairs and Climate distributes datasets via a \gls{gl:wfs} on the platform PDOK \cite{nl2021data}.
The files comprise only the mainland of the Netherlands; overseas territories are not included.
The class precision is very high, encompassing around 320 different plant categories.
So far data is only available for the years 2020 and 2021, of which the file for 2020 was harmonised for EuroCrops.
The datasets fall under the CC0 1.0 licence category which does not impose any limitations to public access.

\paragraph{Portugal}
The Portuguese datasets \cite{pt2021data} are available since 2017, with the file for 2021 harmonised.
Data since 2020 covers the complete national territory of Portugal.
Contrarily, the files for 2017, 2018 and 2019 are split up into regional territories, which had to be merged in a first step.
Moreover, some of the Portuguese regions are missing whereas the national datasets provide a complete and uniform depiction of Portuguese crop cover.
Furthermore, the class precision differs between the regional and the national datasets.
The crop differentiation is moderate for the regional sets with ca. 50 to 150 classes, whereas it is more precise for the national datasets with more than 200 classes.
The files can be accessed via a \gls{gl:wfs} provided by the Portuguese Finance Institute of Agriculture and Fisheries.

\paragraph{Romania} Romania officially does not yet publish crop data but is, according to the \gls{gl:apia}, actively working towards it.
We therefore decided to add a coarse and only regional land cover dataset\cite{ro20xxdata} into EuroCrops in order to give an incentive and an idea of how Romanian data would be integrated in the future.

\paragraph{Slovenia}
The Slovenian dataset \cite{si2021data} covers the territory of the whole country and the years 2019, 2020 and 2021.
The file for 2021 got harmonised.
The class precision is high, with approximately 150 different crop classes.
The files are distributed as shapefiles at the website of the Ministry of Agriculture, Forestry and Food.
Additionally, two text files are published which describe the crop codes assigned to the plants with one file being in Slovenian language, the other one in English.
All data is made publicly available without use restrictions, however, citing the source is required.

\paragraph{Slovakia}
Slovakian data \cite{sk2021data} is available for the years 2020, 2021 and 2022.
The datasets cover all regions of the country.
The file depicting the crop situation in 2021 was harmonised.
The class precision is very high, differentiating between roughly 170 crop types.
The data was sent directly to the project members via e-mail by the Slovakian Agricultural Paying Agency.

\paragraph{Spain}
Spain distributes data under the licence CC BY 4.0 separately for each of its autonomous communities where each one has their own website.
The crop parcel data can be downloaded there as a shapefile in most cases.
So far data has been acquired for the communities of Castile and León, Andalusia and Navarra for the years 2020 and 2021. The Navarra dataset \cite{esna2021data} for 2021 got harmonised.
However, the data is very coarse, differentiating between 21 classes only.

\paragraph{Sweden}
\texttt{GeoJSON} files covering the crop parcels of all of Sweden for the years 2020 and 2021 were sent by a contact person at the Swedish Board of Agriculture\cite{se2021data} to the project members by email.
The files have a medium class precision distinguishing between ca. 80 classes, are published under the CC BY 4.0 licence and data depicting the crop situation in 2021 got harmonised.

%

\section*{Technical Validation}

Regarding the correctness of the underlying original data, it is important to stress that self-declarations build the basis of the input.
From official site, the in-situ controls act as a validation instance to these declarations, but these are just sparse samples and would never be able to cover the entire area.
One approach to actually validate the original data on a bigger scale was introduced by Gounari et al. \cite{gounari2022filtering}, but this would exceed the project tasks.
On our side, we concentrated on a valid harmonisation of the entire dataset.
The validation of the content itself was already discussed in the methods section.\\
In addition to our own approach, the \gls{gl:jrc} of the European Commission is  running validation experiments\cite{schneider2022githubissue7} with their \gls{gl:jrc} MARS database where they compare the areas of the official Eurostats declarations with the ones in EuroCrops.
This is still an ongoing effort and any findings of that will be published together with further EuroCrops versions.


\section*{Usage Notes}

The data is currently published as one shapefile per country on Zenodo\cite{schneider2022ecz} which can, for instance, be opened with \textsc{QGIS}\cite{qgis0000qgis}.
The corresponding mapping\cite{schneider2022githubmappings} files on GitHub are in CSV format with the structure found in Table \ref{tab:mapping}.
In order to use data from a year that has not been harmonised within EuroCrops, it is possible to join the mapping file of a country with the raw vector data file which can be found on the provided national platforms.
By using the correct column in the original dataset, which is indicated in the wiki\cite{schneider2022githubwiki} entry under "Attribute Table" for each country, also other datasets can be harmonised.
This might lead to some missing crop types, as our taxonomy only holds the crop classes occluding in the stated sub-datasets, but we assume that the majority of the crops should be covered.



\section*{Code availability}

As most of the project involved manual work, we did not use a code framework to generate or process the dataset.
All automatic tasks were either translations or sophisticated string matching.
Any code that will help with the improvement of EuroCrops will be published on our GitHub repository\cite{schneider2022github}.


\bibliography{bib/bib}


\section*{Acknowledgements}
The authors and the EuroCrops project receive funding from the German \emph{Federal Ministry for Economic Affairs and Climate Action} on the basis of a resolution of the German Bundestag under reference \texttt{50EE1908} and from the European Union's \emph{Horizon 2020} research and innovation programme under grant agreement No \texttt{101004112}.


\section*{Author contributions statement}

M.S. leads the EuroCrops project, identified data sources, created \gls{gl:hcat}, obtained feedback from authorities and compiled the published shapefiles.
T.S. obtained and documented the individual datasets from the data sources.
F.S. translated the crop classes and analysed the individual datasets.
M.S., T.S. and F.S. verified the crop translations and mappings to \gls{gl:hcat}.
M.K. was involved in the design of the concept and supervised the project.
All authors reviewed the manuscript.


\section*{Competing interests}

The authors declare no competing interests.


\section*{Figures \& Tables}

%
%
%


\begin{table}[ht]
\centering
\begin{tabular}{lp{0.5\textwidth}l}
\hline
\textbf{Country (state)} & \textbf{National agency} & \textbf{Format} \\
\hline
Belgium (Flanders) & \small Department of Agriculture and Fisheries (Departement Landbouw \& Visserij) & shapefile, \texttt{GPKG} \\
Croatia & \small Agency for Payments in Agriculture, Fisheries and Rural Development & \texttt{GPKG} \\
Denmark & \small Danish Agricultural Agency & shapefile \\
Finland & \small Finish Food Authority & \acrshort{gl:wfs} \\
Latvia & \small Rural Support Service Republic of Latvia (Lauku atbalsta dienests) & \acrshort{gl:wfs} \\
Netherlands & \small Ministry of economic affairs and climate (Ministerie van Economische Zaken en Klimaat) via: PDOK platform (Publieke dienstverlening op de kaart) & \acrshort{gl:wfs} \\
Portugal & \small Portuguese Finance Institute of Agriculture and Fisheries (Instituto de Financiamento da Agricultura e Pescas) & \acrshort{gl:wfs} \\
Slovenia & \small Ministry of Agriculture, Forestry and Food (Ministrstvo za Kmetijstvo, Gozdarstvo in Prehrano) & shapefile \\
\hline
\end{tabular}
\caption{\label{tab:ministries}
    \textbf{National (ministry) website}:
    The majority of the EuroCrops data sources were websites, usually hosted by the respective ministry or agency.
    These websites are usually in the national language, without any English translation which makes the discoverability and accessibility of the data laborious for international researchers.
}
\end{table}

\begin{table}[ht]
    \centering
    \begin{tabular}{lll}
        \hline
        \textbf{Country (state)} & \textbf{Portal} & \textbf{Format} \\
        \hline
        Austria & data.gv.at & \texttt{GPKG} \\
        Belgium (Wallonia) & Géoportail de la Wallonie & shapefile \\
        France & data.gouv.fr & shapefile \\
        Germany (North Rhine-Westphalia) & OpenGeodata.NRW & shapefile \\
        Germany (Lower Saxony) & Landentwicklung und Agrarförderung Niedersachsen-Portal & shapefile \\
        Lithuania & geoportal.lt & shapefile \\
        Spain & "Sicpac" portal for each Autonomous Community & shapefile, \acrshort{gl:wfs} \\
        \hline
    \end{tabular}
    \caption{\label{tab:geoportal}
        \textbf{National geoportal}:
        Some countries or regions actively participate in Europe's open data initiative and publish their crop data on a national geoportal.
        The goal of these portals is to make data available to the public sector and lower the entry barrier to letting citizens actively participate.
}
\end{table}

\begin{table}[ht]
    \centering
    \begin{tabular}{ll}
        \hline
        \textbf{Country} & \textbf{Format} \\
        \hline
        Slovakia & shapefile \\
        Sweden & shapefile \\
        \hline
    \end{tabular}
    \caption{\label{tab:direct}
        \textbf{Direct contact}:
        Data from Slovakia and Sweden was directly sent to the project members by a contact person in the respective country.}
\end{table}

\begin{table}[ht]
    \centering
    \begin{tabular}{lll}
        \hline
        \textbf{Country} & \textbf{Portal} & \textbf{Format} \\
        \hline
        [Austria & data.europa.eu & \texttt{GPKG}] \\
        Estonia & INSPIRE Geoportal & \acrshort{gl:wfs} \\
        Romania & INSPIRE Geoportal & shapefile \\
        \hline
    \end{tabular}
    \caption{\label{tab:international} \textbf{International Platform}
        This table lists all the countries which data was acquired through an international platform.
        The data for Austria would be available via "data.europa.eu", but for EuroCrops we downloaded it directly from the national website.
}
\end{table}

\begin{table}[ht]
    \centering
    \scalebox{0.8} {
    \begin{tabular}{llllllllr}
        \hline
            \small \textbf{ID} & \small \textbf{FLIK} & \small \textbf{AREA\_HA} & \small \textbf{COD}E & \small \textbf{CODE\_TXT} & \small \textbf{USE\_CODE} & \small \textbf{USE\_TXT} & \small \textbf{WJ} & \small \textbf{DAT\_BEARB} \\
            \hline
            \small 4597509	& \small DENWLI0543050566 & 2.1808 & 411 & \small Silomais (als Hauptfutter) & AF & \small Ackerfutter & 2021 & 2021/03/01 \\
            \small 4597510	& \small DENWLI0543051616 & 1.5319 & 459 & \small Grünland (Dauergrünland) & GL & \small Dauergrünland & 2021 & 2021/03/01 \\
            \small 4597641 & \small DENWLI0542022516 & 1.0293 & 480 & \small Streuobst mit DGL-Nutzung & GL & \small Dauergrünland & 2021 & 2021/03/02 \\
            \small 4597657 & \small DENWLI0541093620 & 2.4966 & 459 & \small Grünland (Dauergrünland) & GL & \small Dauergrünland & 2021 & 2021/03/02 \\
            \small 4597810 & \small DENWLI0540163053 & 1.162 & 121 & \small Winterroggen & GT & Getreide & 2021 & 2021/03/04 \\
        \hline
\end{tabular}
}
\caption{\label{tab:raw_data}
    After downloading the respective national raw datasets, we first examined the attribute tables and extracted the values of the columns representing the cultivated crops.
    In the given example, showing raw data from North Rhine-Westphalia (Germany) \cite{denrw2021data}, multiple columns representing the cultivated crops can be determined.
    Thus a selection had to be made.
    In this case, values from the "CODE\_TXT" column were translated and matched with the occurring classes in \gls{gl:hcat}.
    Each time we discovered a class that was not represented in the taxonomy yet, we included it and started the harmonisation process again.
    The corresponding vector data to this file is illustrated in Figure \ref{fig:nrw} and the same attribute table enriched with the EuroCrops columns is shown in Table \ref{tab:ec_data}.
}
\end{table}

\begin{table}[ht]
    \centering
    \scalebox{0.65} {
        \begin{tabular}{llllllrllr}
            \hline
            \small \textbf{ID} & \small \textbf{...} & \small \textbf{CODE\_TXT} & \small \textbf{USE\_CODE} & \small \textbf{USE\_TXT} & \small \textbf{WJ} & \small \textbf{DAT\_BEARB} & \small \textbf{EC\_trans\_n} & \small \textbf{EC\_hcat\_n} & \small \textbf{EC\_hcat\_c} \\
            \hline
            \small 4597509 & ... & \small Silomais (als Hauptfutter) & AF & \small Ackerfutter & 2021 & 2021/03/01 & \small Silage maize (as staple feed) & green\_silo\_maize & 3301090400 \\
            \small 4597510	&... & \small Grünland (Dauergrünland) & GL & \small Dauergrünland & 2021 & 2021/03/01 & \small Grassland (permanent grassland) & pasture\_meadow\_grassland\_grass & 3302000000 \\
            \small 4597641 &... & \small Streuobst mit DGL-Nutzung & GL & \small Dauergrünland & 2021 & 2021/03/02 & \small Orchards with Permanent grassland use & orchards\_fruits & 3303010000 \\
            \small 4597657 &... & \small Grünland (Dauergrünland) & GL & \small Dauergrünland & 2021 & 2021/03/02 & \small Grassland (permanent grassland) & pasture\_meadow\_grassland\_grass & 3302000000\\
            \small 4597810 &... & \small Winterroggen & GT & Getreide & 2021 & 2021/03/04  & \small Winter rye & winter\_rye & 3301010301 \\
            \hline
        \end{tabular}
    }
    \caption{\label{tab:ec_data}
        This table shows an example of a final EuroCrops data attribute table.
        While the original columns as shown in Table \ref{tab:raw_data} remains the same ("FLIK", "AREA\_HA" and "CODE" have been abbreviated in this print), three additional attributes were added:
        Firstly, "EC\_trans\_n" is the direct translation of the crop name in its original language.
        Then, correspondingly, "EC\_hcat\_n" is the matched name of that particular crop in \emph{HCAT}.
        These names are all lowercase and with underscores to make them easier to  process automatically.
        Lastly, the column "EC\_hcat\_c" shows the \gls{gl:hcat} code that puts the \gls{gl:hcat} names into a hierarchical structure.
        A more detailed explanation of \gls{gl:hcat} is presented in the publication by Schneider et al.\cite{schneider2021epe}.
    }
\end{table}

\begin{table}[ht]
    \centering
    \scalebox{0.9} {
    \begin{tabular}{lllll}
        \hline
        \textbf{original\_code} & \textbf{original\_name} & \textbf{translated\_name} & \textbf{HCAT2\_name} & \textbf{HCAT2\_code} \\
        \hline
        459 & Grünland (Dauergrünland) & Grassland (permanent grassland) & pasture\_meadow\_grassland\_grass & 3302000000 \\
        411 & Silomais (als Hauptfutter) & Silage maize (as staple feed) & green\_silo\_maize & 3301090400 \\
        \hline
    \end{tabular}
}
    \caption{\label{tab:mapping}
        In our aforementioned GitHub repository, we publish for each country a so-called mapping file.
        This file contains the set of occurring crops in one original file, together with its translation and the corresponding \gls{gl:hcat} name and code.
        Note, that even though it says {HCAT2} in the column names, it is the same as the previously mentioned \gls{gl:hcat}.
        As the initial, prototyped taxonomy is not used any more.
        The shown example is an extraction of the sub-dataset which is already presented in Figure \ref{fig:nrw} and Tables \ref{tab:raw_data} and \ref{tab:ec_data}.
        The entire file is available on GitHub and could for example be used to translate and map a dataset from North Rhine-Westphalia (Germany) \cite{denrw2021data} of another year than 2021.
    }
\end{table}


\end{document}